# High-precision Beam Optics Calculation of the HIAF-BRing Using Measured Fields

Ke Wang [a, b, c, d], Li-Na Sheng [b, c], Geng Wang [b, c], Wei-Ping Chai [b, c], You-Jin Yuan [b, c], Jian-Cheng Yang [b, c], Guo-Dong Shen [b, c*], Liang Lu [a, d†]

a. Sino-French Institute of Nuclear Engineering and Technology, Sun Yat-sen University, Zhuhai 519082, P. R. China

b. Institute of Modern Physics, Chinese Academy of Sciences, Lanzhou 730000, P. R. China

c. University of Chinese Academy of Sciences, Beijing 100049, P. R. China

d. United Laboratory of Frontier Radiotherapy Technology of Sun Yat-sen University & Chinese Academy of Sciences Ion Medical Technology Co., Ltd, Guangzhou, P. R. China

Abstract

The construction of the High Intensity heavy ion Accelerator Facility (HIAF) has been completed, with current efforts focused on subsystem commissioning. Beam commissioning is scheduled for autumn 2025, marking a critical milestone in the HIAF project. This paper presents high-precision optics calculations for the Booster Ring (BRing) of HIAF, a key component for achieving stable heavy-ion beam acceleration. Leveraging high-precision magnetic field data, each magnet is divided into hundreds of slices, thus establishing a high-precision sliced optics model for BRing. Detailed calculations of BRing's optics are presented in this work. Critical parameters including tunes and betatron functions of the lattice based on the measured magnetic fields and those of the ideal lattice have been compared. The results highlight the impact of realistic magnetic field on beam dynamics and provide essential insights for accelerator tuning and optimization. These findings serve as a fundamental reference for beam commissioning and long-term operation, ensuring beam stability and performance reproducibility in HIAF.

Keywords: Lattice, Accelerator tuning, Magnet sliced model, Magnetic field measurement

## 1. Introduction

The High Intensity heavy ion Accelerator Facility (HIAF) [1-3] is designed and constructed by the Institute of Modern Physics, Chinese Academy of Sciences. It is mainly composed of the superconducting electron cyclotron resonance ion source (SECR), the superconducting linear accelerator (iLinac), the Booster Ring (BRing) [4], the High energy FRagment Separator (HFRS) [5, 6], and the Spectrometer Ring (SRing) [7].

* Corresponding author: Guo-Dong Shen, Email: shenguodong@impcas.ac.cn.

† Corresponding author: Liang Lu, Email: luliang3@mail.sysu.edu.cn.

Its layout is shown in Figure 1. The construction of HIAF has been essentially completed. Currently, the commissioning work of sub-systems (such as magnets, power supplies, etc.) is in progress. The beam commissioning is scheduled for this autumn. Structurally, BRing comprises three arc sections and three straight sections. Each arc section contains 16 dipole magnets and 16 quadrupole magnets, while each straight section houses 10 quadrupole magnets. BRing is capable of accelerating $^{238}U^{35+}$ ions to 0.8 GeV/u with a beam intensity of 1.0E+11 ppp (particles per pulse). Both energy and beam intensity parameters significantly exceed those of China's existing Heavy Ion Research Facility in Lanzhou-Cooler Storage Ring (HIRFL-CSR) [8]. In particular, the beam intensity will reach a world-leading level, thereby imposing substantially enhanced requirements for accelerator tuning.

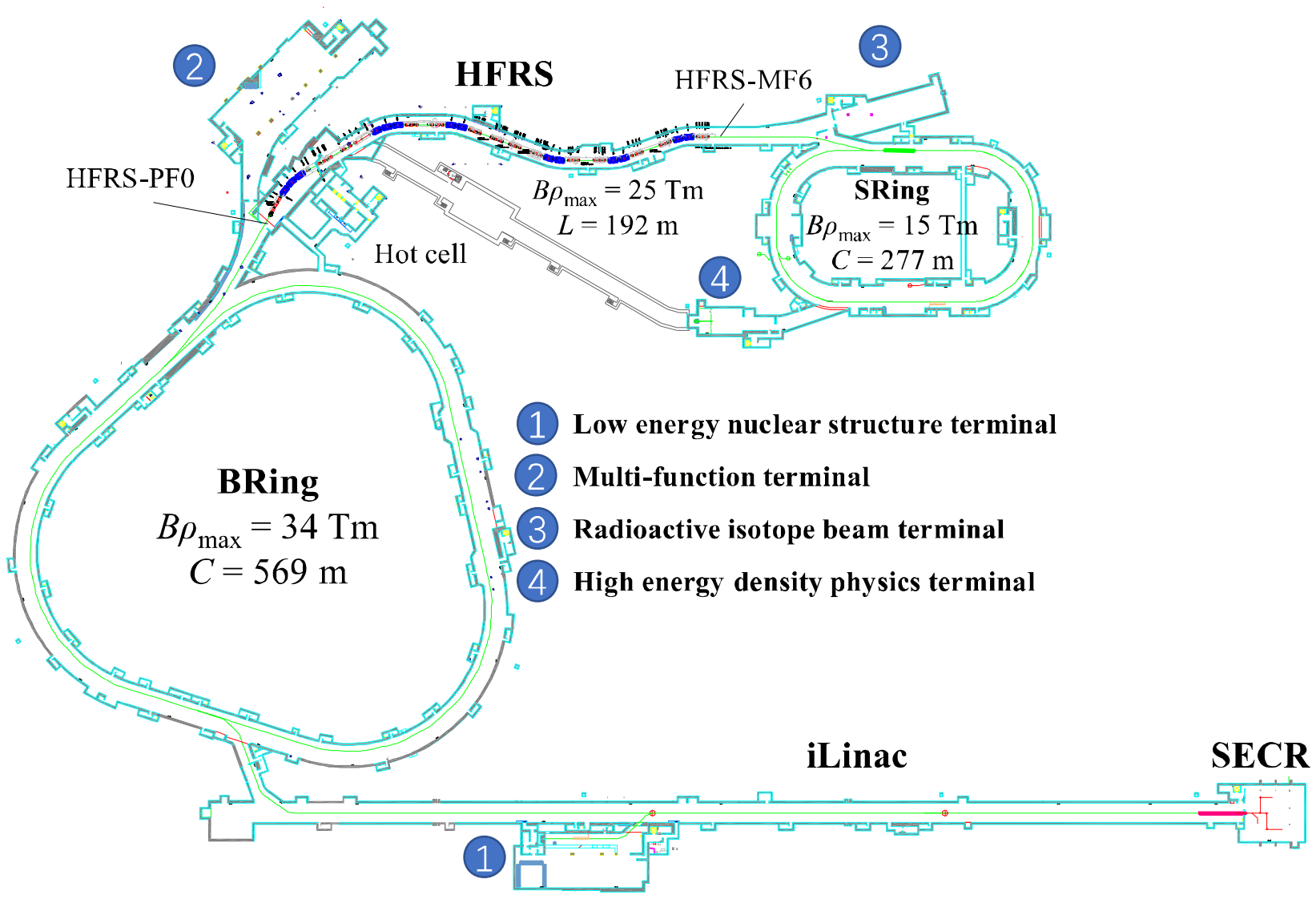

Figure 1. Layout of the HIAF.

Table 1. Design lengths and effective lengths of the magnets in BRing under low (L), medium (M), and high (H) field.

| | Dipole | QL850D180 | QL1000D180 | QL1400D260 |
|---|---|---|---|---|
| Design length [mm] | 2814.34 | 850 | 1000 | 1400 |
| Effective Length [mm] | 2824.37(L) | 872.25(L) | 1011.34(L) | 1407.48(L) |
| | 2835.34(M) | 870.05(M) | 1010.94(M) | 1407.97(M) |
| | 2812.37(H) | 858.67(H) | 1001.85(H) | 1400.24(H) |

The beam tuning of HIAF is based on MAD-X [9]. In the ideal lattice optics model, magnet lengths are set to their design lengths. In terms of fringe field, dipole magnets incorporate fringe field effects using MAD-X's built-in fringe field integrals[10], while quadrupole magnets adopt the hard-edge approximation[11]. And without considering the overlap of magnetic fields between magnets and the longitudinal homogeneity of magnetic fields. The calculated focusing strength of the quadrupole magnet in this case should be inappropriate. Therefore, it is necessary to establish high-precision magnet models in the optics model. Figure 2 illustrates the magnetic flux density distribution along the z-axis (x=0, y=0) for BRing's dipole and quadrupole magnets. Effective lengths of magnets are calculated from magnetic field measurement data, as summarized in Table 1. The results indicate that both dipole and quadrupole magnets in BRing exhibit effective lengths exceeding their design specifications in most case.

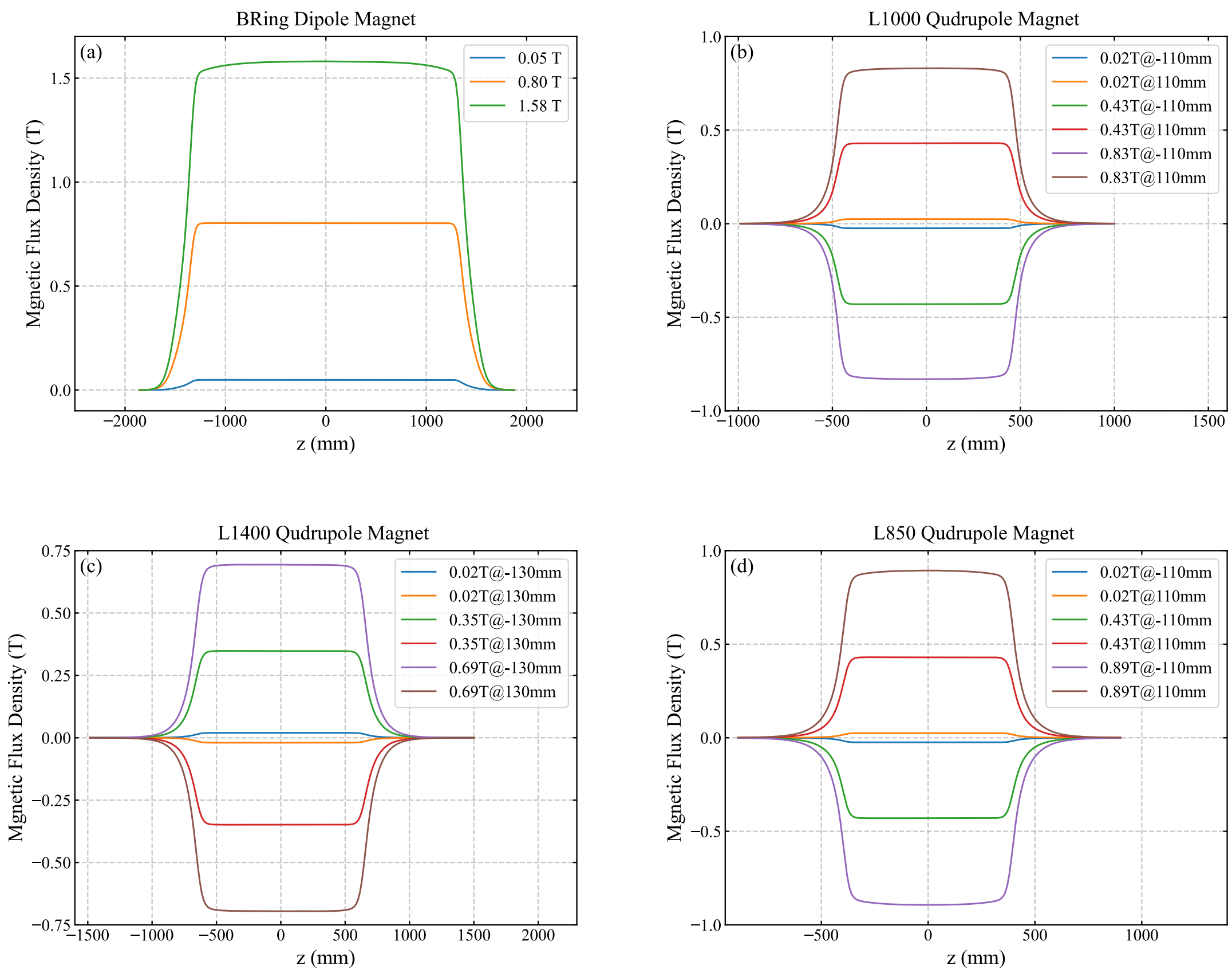

Fig. 2. The magnetic fields of dipole magnet and quadrupole magnets in BRing.

Based on the high-precision measured magnetic field data of the dipole magnets and quadrupole magnets in BRing, this paper establishes a high-precision optics model of BRing to achieve accurate beam tuning. High-precision optics calculation is crucial for the stable operation of BRing and any large accelerator.

## 2. High-precision modeling

Table 2. Magnet modeling details.

| | Dipole | QL850 | QL1000 | QL1400 |
|---|---|---|---|---|
| Length of Slice [mm] | 18.762 | 10 | 10 | 15 |
| Number of Measurement Points | 201 | 183 | 201 | 201 |
| Number of Slices | 200 | 182 | 200 | 200 |
| Total length [mm] | 3752.458 | 1820 | 2000 | 3000 |

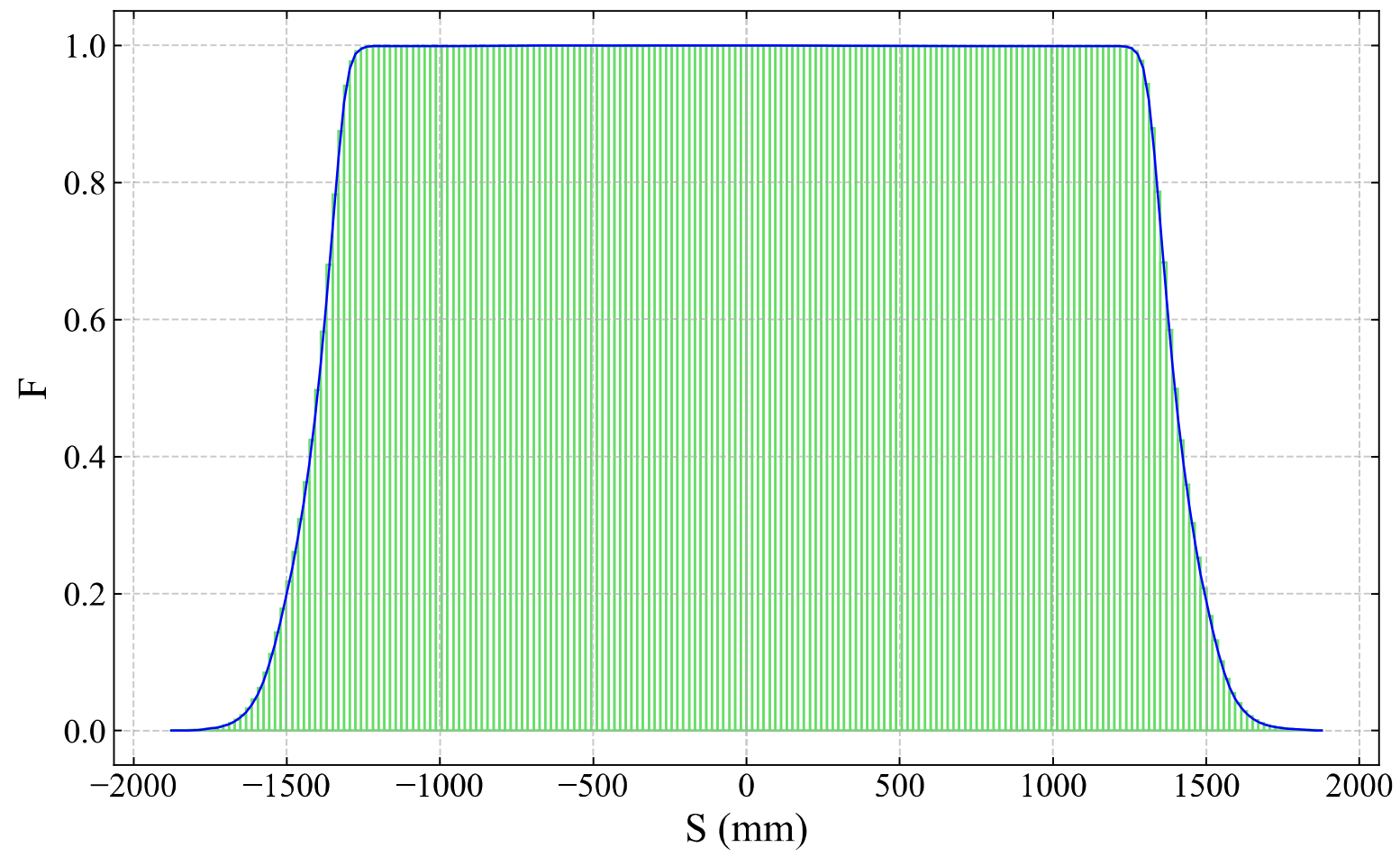

Fig. 3. Schematic of the sliced model of the dipole magnet (0.8 T) in BRing. The width of the red rectangle represents the length of the slice, and the height represents the corresponding normalized magnetic flux intensity $F_i$ of the i-th slice.

The bending angle $Angle_i$ of each slice of the dipole magnet in BRing is calculated via Equation (1), where $L_s$ is the length of the slice, L is the design length of the magnet, $L_{eff}$ is the effective length of the magnet, R is the bending radius, and $F_i$ is the corresponding normalized magnetic flux intensity of the i-th slice. Obviously, the integral of F along S is equal to the effective length of the magnet, the sum of all $Angle_i$ of the slices is equal to the design bending angle of the magnet, and the sum of the lengths is equal to the length of the measured magnetic field. In this paper, $L/L_{eff}$ is called the correction factor. As shown in Table 2, based on the number of measurement points from the point measurement of magnetic field, the dipole magnet is divided into 200 slices. Each slice has a length of 18.762 mm, and the total length is equal to the length of an arc with a radius of 21.5 m and a central angle of 10 degrees. As depicted in Fig. 3, the width of the green rectangle represents the length of each slice, and the height corresponds to the normalized magnetic flux intensity F of that slice. Meanwhile, the fringe field integrals at entrance and exit of the dipoles have been recalculated.

$$\mathrm{Angle}_i = \frac{L_s}{R/F_i} \cdot \frac{L}{L_{eff}} . \tag{1}$$

$$\sum \mathrm{Angle}_i = \mathrm{Angle} . \tag{2}$$

$$L_s \sum F_i = L_{eff} . \tag{3}$$

For the quadrupole magnet, the focusing strength K of each slice is calculated via Equation (4). The total length of all slices of a quadrupole magnet is equal to the length of the measured magnetic field. The QL850 quadrupole magnet is divided into 182 slices, with each slice having a length of 10 mm, and the total length being 1820 mm. The QL1000 quadrupole magnet is divided into 200 slices, with each slice having a length of 10 mm, and the total length being 2000 mm. The QL1400 quadrupole magnet is divided into 200 slices, with each slice having a length of 15 mm, and the total length being 3000 mm. This modeling method takes into account the effective length of the quadrupole magnet. Since the effective length of the quadrupole magnet is longer, its original focusing strength is no longer appropriate (it is excessive), and the corresponding excitation current is also inappropriate. A new K value can be calculated through "matching", and then a new excitation current can be calculated according to the I-GL curve of the magnet.

$$K_i = K \cdot F_i \, . \tag{4}$$

$$L_s \sum K_i = K L_{eff} \, . \tag{5}$$

The second modeling method for the quadrupole magnet is to multiply the focusing strength of each slice by the correction factor, as shown in Equation (6), following the same methodology applied to the dipole magnet. This approach ensures that the longitudinal field integral of the magnet remains unchanged before and after slicing while providing greater convenience for calculating the GL of the quadrupole magnet. Consequently, this method has been adopted in practical applications.

$$K_i = K \cdot F_i \cdot \frac{L}{L_{eff}} \, . \tag{6}$$

$$L_s \sum K_i = KL \, . \tag{7}$$

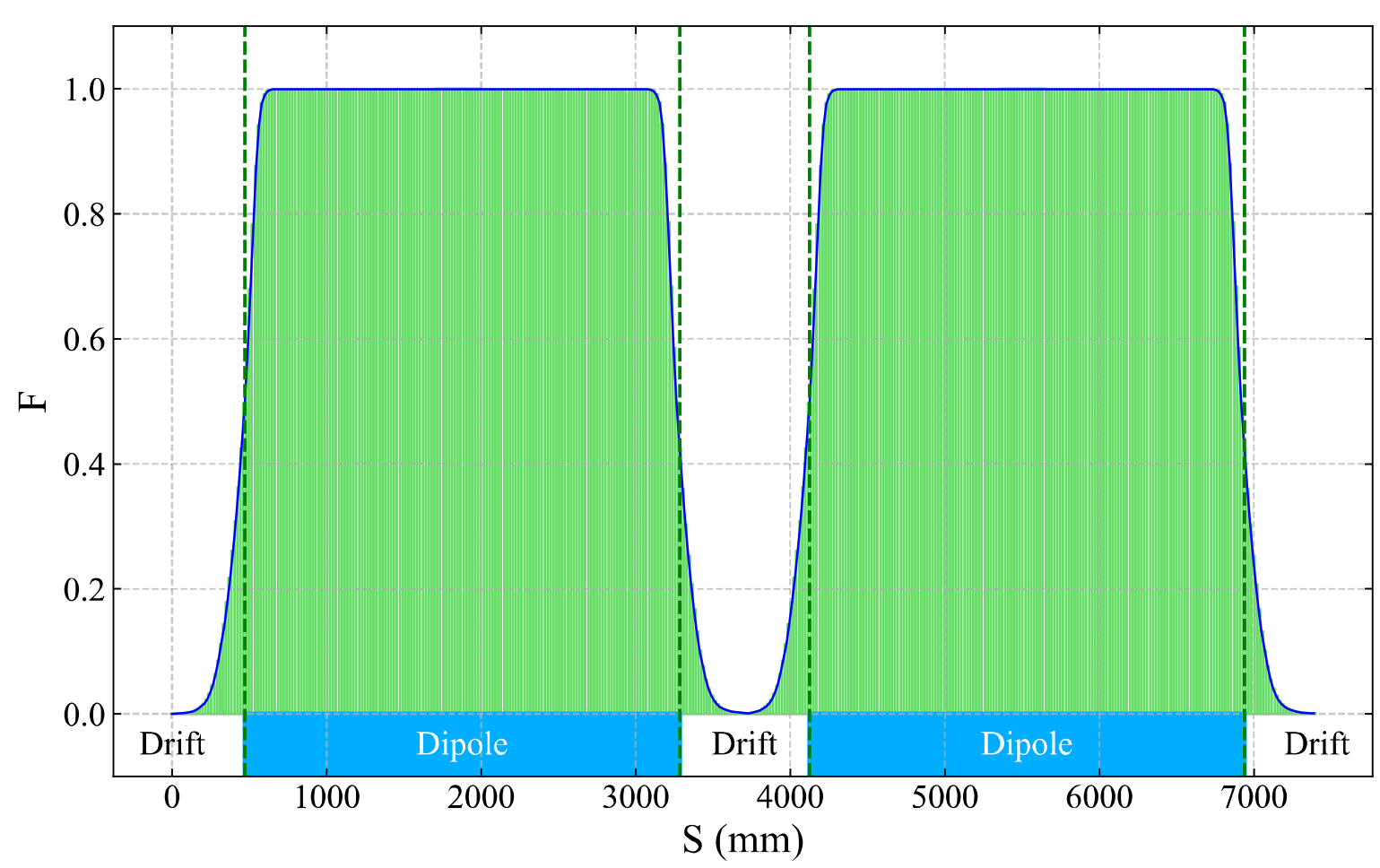

Fig. 4. Combination of two dipole magnets and three adjacent drifts.

The length of the drift between the dipole magnets of BRing is 840 mm. A 98.114 mm longitudinal region at the drift center forms an overlap zone where the magnetic fields from both left and right dipole magnets coexist. The sliced model takes into account the superposition of the magnetic fields between the magnets, as shown in Fig. 4. For the dipole magnets, the slices of two magnets are combined and packaged as a “line”. Through this modeling method, not only the fringe fields of the magnets, but also the superposition of magnetic fields between adjacent magnets and the longitudinal variation of the magnetic field are taken into consideration. Due to the good consistency of the magnets, the point measurement data of one magnet from each type is selected for modeling.

## 3. Optics calculations and discussion

The betatron functions (β), horizontal dispersion functions (DX), tunes, and chromaticity of the ring are calculated using MAD-X. Optics models include:

- the original ideal model,
- the dipole-only sliced model (D),
- the quadrupole-only sliced model without correction factor (Q),
- the quadrupole-only sliced model with correction factor (QC),
- the all-magnets sliced model without correction factor (A),
- the all-magnets sliced model with correction factor (AC).

Fig. 5 shows the β and horizontal dispersion functions of BRing's ideal optics. Fig. 6 shows the differences between sliced optics without matching and ideal optics in terms of β and horizontal dispersion functions. Fig. 7 shows the changes in each quadrupole magnet after matching. Fig. 8 shows the differences between sliced-optics with matching and ideal optics in terms of β and horizontal dispersion functions.

In the standard operation mode, the horizontal tune of BRing is 9.470, and the vertical tune is 9.430, as shown in Table 3. When only the dipole magnets are sliced, the horizontal tune changes to 9.461, and the vertical tune changes to 9.432. Overall, the horizontal focusing is somewhat weakened, while the vertical focusing is slightly enhanced. It can be seen that under realistic magnetic field conditions, although the integral of the magnetic field of the dipole magnet along the reference orbit is maintained unchanged, both the horizontal and vertical tunes have undergone changes to varying degrees, with the reduction in the horizontal tune being particularly significant. This is due to the fringe field and the homogeneity of the magnetic field in the longitudinal direction, so it is also necessary to adjust the quadrupole magnets synchronously. The β functions hardly changes, while the horizontal dispersion functions in the straight section is no longer constantly 0, but fluctuates slightly around 0.

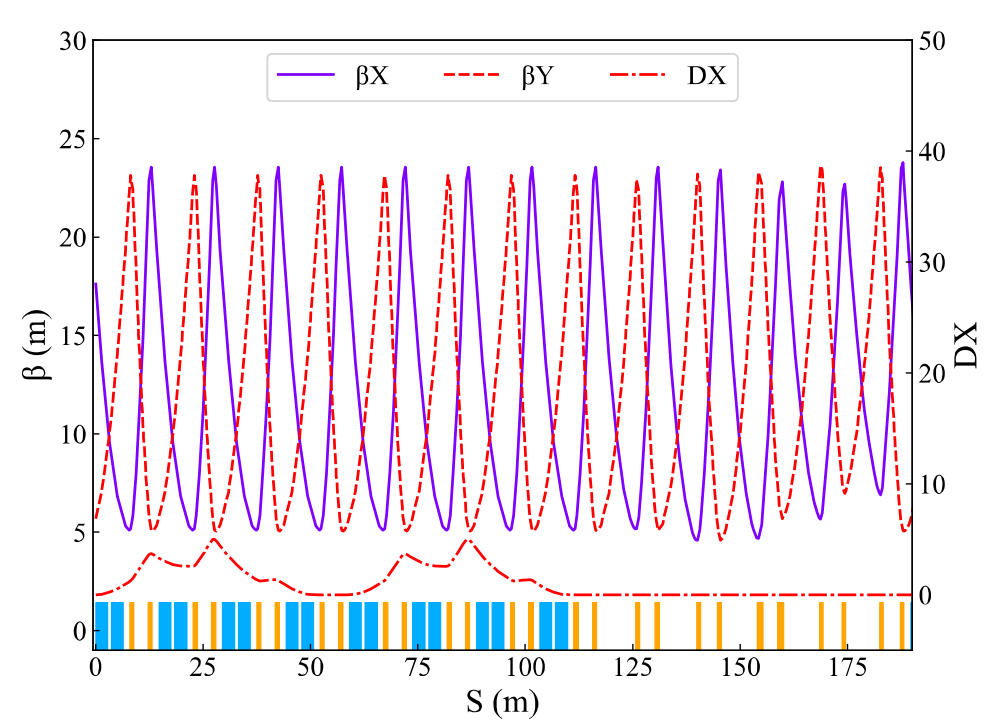

Fig.5. β and horizontal dispersion functions of BRing's ideal optics (1/3 circumference). Blue squares represent dipole magnets, and yellow squares represent quadrupole magnets.

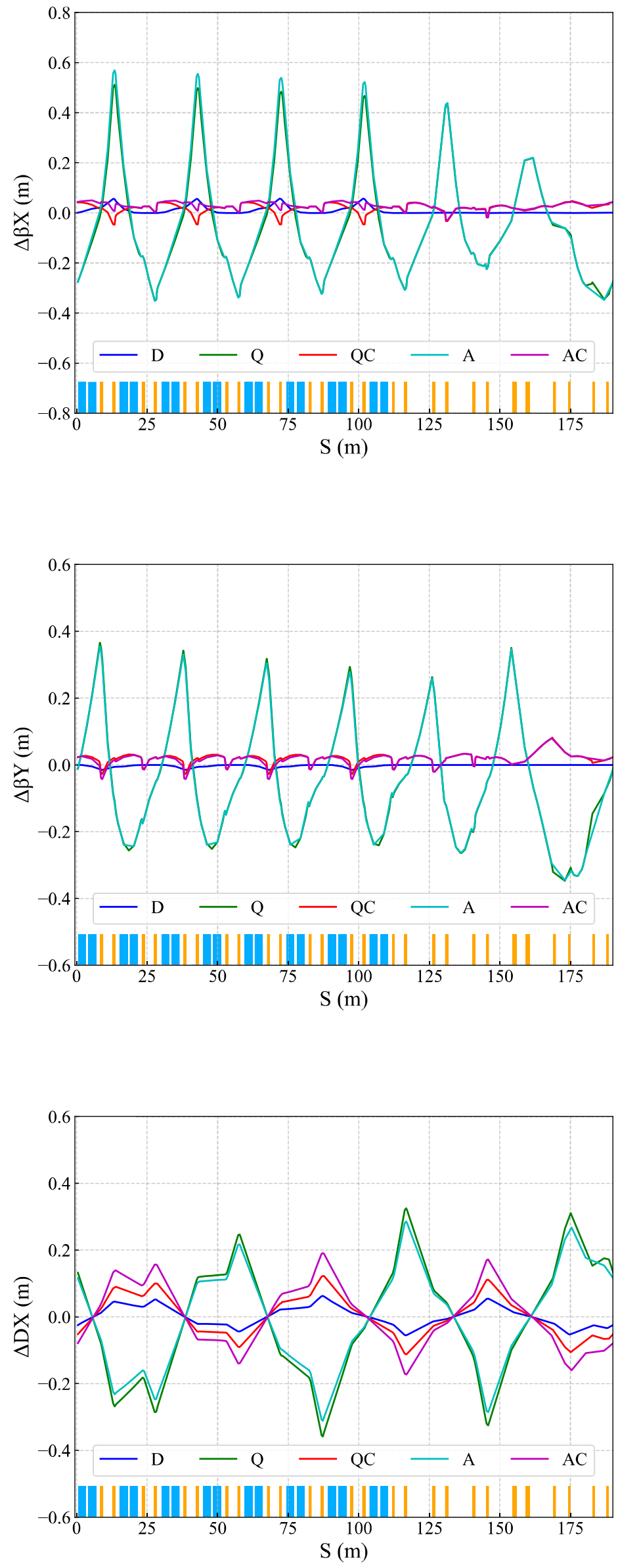

Fig. 6. Differences of β and horizontal dispersion functions (1/3 circumference) without matching.

When only the quadrupole magnets are sliced, without using the correction factor, the horizontal tune increases by 0.081, and the vertical tune increases by 0.076, indicating that both the horizontal and vertical focusing are enhanced. This is because the effective length of the quadrupole magnet is longer than the design length, and the longitudinal integral of magnetic field is slightly greater than what is required. The β functions shows a slight increase, and the fluctuation of the horizontal dispersion functions in the straight section becomes more significant, varying between -0.4 m and 0.4 m. When the correction factor is used, both the horizontal and vertical tunes decrease significantly compared to the case without using the correction factor, and they are smaller than those in the ideal optics. The focusing in both horizontal and vertical directions is weakened although the integral of the magnetic field of the quadrupole magnet along the reference orbit is kept unchanged. The increase in the β functions is very small, and the horizonal dispersion functions in the straight section varies between -0.1 m and 0.1 m.

When both the dipole magnets and the quadrupole magnets are sliced, and the correction factor is not used for the quadrupole magnets, the horizontal tune increase by 0.093, and the vertical tune increase by 0.100. The β functions shows a slight increase, and the horizonal dispersion functions in the straight section varies between -0.3 m and 0.3 m. When the correction factor is used for the quadrupole magnets, both the horizontal and the vertical tunes decrease significantly compared to the situation without using the correction factor, and they are smaller than those in the ideal model. The increase in the β functions is very small, and the horizonal dispersion functions in the straight section varies between -0.2 m and 0.2 m.

Table 3. Tunes, chromaticity, and the maximum of β functions.

| | O | D | Q | QC | A | AC |
|---|---|---|---|---|---|---|
| $Q_x$ | 9.470 | 9.461 | 9.551 | 9.443 | 9.563 | 9.434 |
| $Q_y$ | 9.430 | 9.432 | 9.506 | 9.403 | 9.530 | 9.406 |
| $dQ_x$ | -11.725 | -11.717 | -11.892 | -11.671 | -11.914 | -11.662 |
| $dQ_y$ | -11.355 | -11.335 | -11.515 | -11.303 | -11.525 | -11.282 |
| $\beta x_{max}$/m | 23.875 | 23.876 | 24.035 | 23.901 | 24.200 | 23.903 |
| $\beta y_{max}$/m | 23.738 | 23.738 | 23.905 | 23.802 | 23.842 | 23.802 |

Table 4. The focusing strengths [$m^{-2}$] of the quadrupole magnets.

| | O | D | Q | QC | A | AC |
|---|---|---|---|---|---|---|
| QD1000 | -0.21755340 | -0.21716492 | -0.21599049 | -0.21820261 | -0.21564468 | -0.21787888 |
| QF1000 | 0.22215098 | 0.22225811 | 0.22035319 | 0.22272129 | 0.22046977 | 0.22284616 |
| QD1400 | -0.15395567 | -0.15538188 | -0.15192865 | -0.15385694 | -0.15348675 | -0.15532827 |
| QF1400 | 0.15381942 | 0.15275498 | 0.15282367 | 0.15405681 | 0.15181609 | 0.15297200 |
| QD850A | -0.21383245 | -0.21379166 | -0.21093398 | -0.21402387 | -0.20956120 | -0.21331502 |
| QF850A | 0.21735381 | 0.21760909 | 0.21525186 | 0.21762679 | 0.21458014 | 0.21770503 |
| QD850B | -0.23415291 | -0.23621169 | -0.23106243 | -0.23414606 | -0.23367080 | -0.23603996 |
| QF850B | 0.22631246 | 0.22840308 | 0.22298794 | 0.22662601 | 0.22572207 | 0.22869130 |

Matching with the tunes and chromaticity as the objectives, the new focusing strengths of the quadrupole magnets can be obtained, as shown in Table 4. Fig. 7 shows the variation of the focusing strengths of various types of quadrupole magnets. It can be seen that in the case of QC, the variation range of the focusing strength is the smallest. In the three cases where the dipole magnets are sliced (D, A, and AC), the focusing strengths of some quadrupole magnets change significantly. This is mainly due to the fringe field of the dipole magnet. When only the quadrupole magnets are sliced without correction factor, the focusing strengths of all quadrupole magnets decrease to varying degrees. This should be because the effective length of the magnets is longer than the design length, and the focusing strength needs to be reduced. After matching, the fluctuations of the horizontal dispersion functions have become very small in all cases, and they are basically consistent with those of the ideal optics, as shown in Fig. 8. Minor changes in the β functions are practically inevitable.

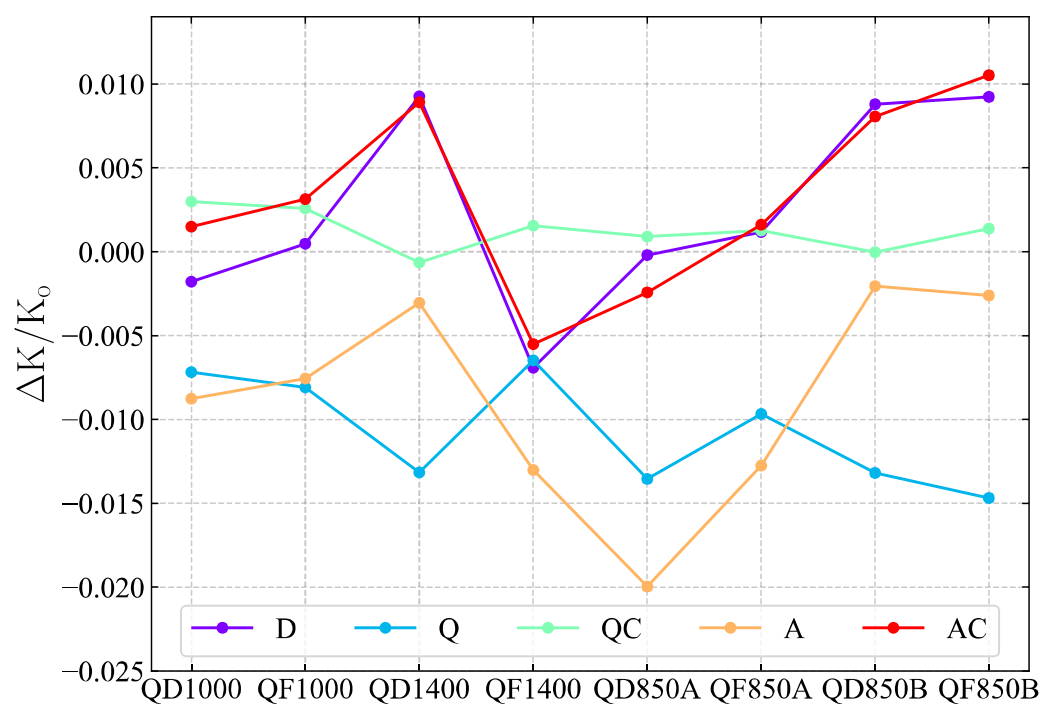

Fig. 7. Variation of the focusing strengths of the quadrupole magnets.

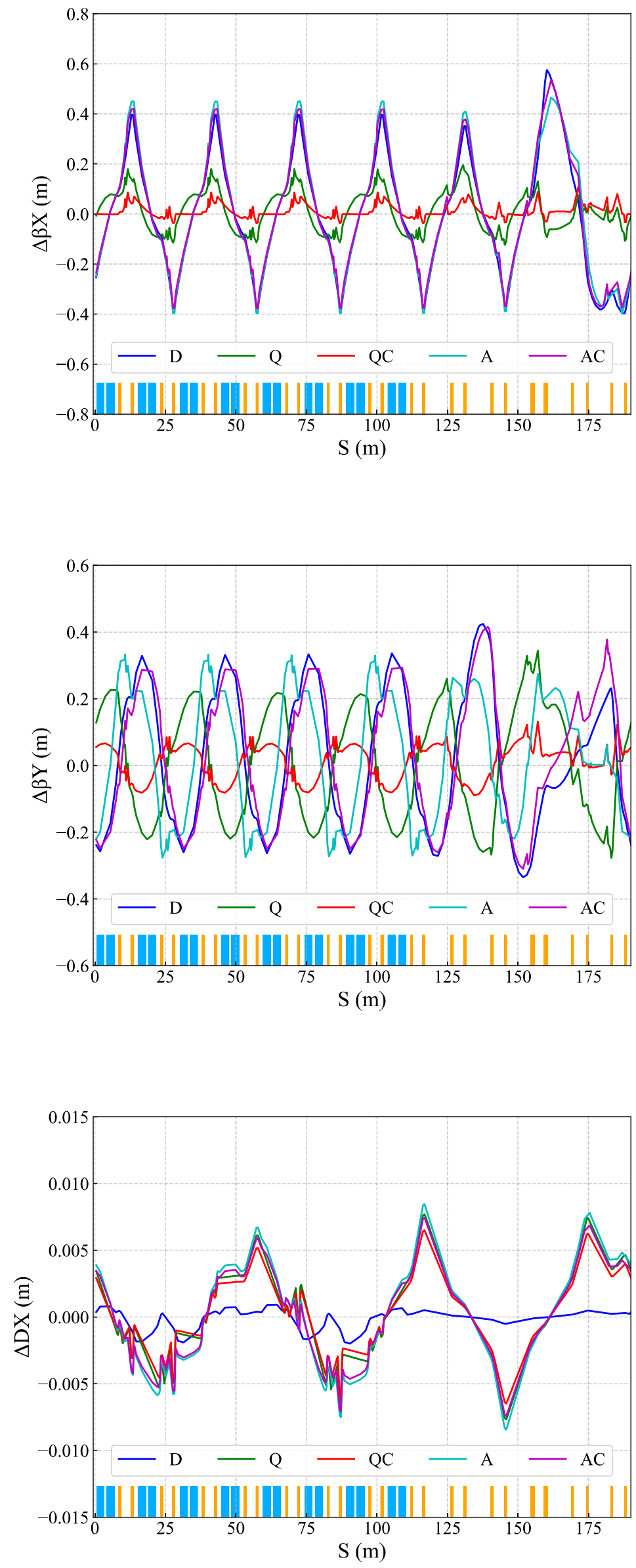

Fig. 8. Differences of β and horizontal dispersion functions with matching (1/3 circumference).

## 4. Summary

This study focuses on the high-precision beam optics calculations of HIAF-BRing and conducts a detailed analysis of the influences of the realistic magnetic fields of dipole magnets and quadrupole magnets on the working points, β functions, horizontal dispersion functions, and so on. According to the magnetic field point measurement data of the dipole magnets and quadrupole magnets, the magnets are longitudinally subdivided into hundreds of segmentations, so as to simulate the fringe field effect and longitudinal homogeneity of the real magnetic field. For the dipole magnets, the superposition of the magnetic fields between the magnets is also taken into account. The simulation results show that the fringe fields of the dipole magnets weak the horizontal focusing but enhances the vertical focusing to some extent, while the fringe fields of the quadrupole magnets weak both the horizontal and vertical focusing. Under the condition of maintaining constant longitudinal field integral, the dipole magnet's field reduces the horizontal tune by approximately 0.009 while increasing the vertical tune by about 0.002. Meanwhile, the quadrupole magnet's field decreases both the horizontal and vertical tunes by approximately 0.027 each. These findings show that a high-precision optics based on the measured magnetic field is necessary for the precise tuning of the accelerator. Finally, through MAD-X simulation matching, the stable control of the tunes, β functions and horizontal dispersion functions of the whole ring is achieved. This study provides key technical support for the commissioning and stable operation of HIAF-BRing, and offers an important reference for the physics design optimization of complex accelerators in the future. Further research, including how the sliced model affects geodesic alignment, dynamic aperture, and other related aspects, will be conducted in subsequent work to provide a more comprehensive analysis of the machine performance.